\documentclass[conference, 10pt]{IEEEtran}
\usepackage{cite}
\usepackage{amsmath,amssymb,amsfonts,bm}
\usepackage{algorithmic}
\usepackage{graphicx}
\usepackage{textcomp}
\usepackage{xcolor}
\usepackage{algorithm} 
\usepackage{booktabs} 
\usepackage{enumitem}  
\usepackage{epstopdf}
\usepackage{multirow}
\usepackage{array}
\usepackage{amsthm}
\usepackage{lipsum}
\usepackage{stfloats}
\usepackage{subfigure}
\usepackage{cases}
\usepackage[OT1]{fontenc}
\usepackage[utf8]{inputenc}

\title{Antenna Coding and Digital Precoding for Limited Feedback MIMO Systems Using Pixel Antennas}
\author{ \IEEEauthorblockN{Zhetong Li$^1$ and Hongyu Li$^2$}

\IEEEauthorblockA{$^1$ School of Information and Electronics, Beijing Institute of Technology, Beijing, China}

\IEEEauthorblockA{$^2$ Internet of Things Thrust, The Hong Kong University of Science and Technology (Guangzhou), Guangzhou, China \\ Corresponding Author E-mail: \texttt{hongyuli@hkust-gz.edu.cn}}
}

\begin{document}
\maketitle

\begin{abstract}
Pixel antennas enable antenna coding, a technique that can provide more degrees of freedom in wave manipulation, to enhance wireless communications.
However, acquiring full channel state information (CSI) at the transmitter incurs prohibitive overhead due to the unique hardware constraints from pixel antennas.
This paper thus proposes a limited feedback multi-input multi-output (MIMO) system using pixel antennas, where the antenna coder and digital precoder are designed based on pre-defined codebooks and efficient index feedbacks. 
We first derive the optimal digital precoder under practical power constraints that provides insights on simplifying the joint codebook construction for antenna coder and digital precoder.
We then develop a low-complexity offline codebook construction algorithm that enables subsequent codebook designs for the antenna coder and digital precoder.
Simulation results demonstrate that the proposed scheme significantly outperforms unconstrained MIMO systems using conventional antennas with fixed configurations.    
\end{abstract}


\section{Introduction}
The transition to sixth-generation (6G) wireless networks demands unprecedented capabilities with terabit-per-second data rates, ultra-low latency, and massive connectivity. To meet these stringent requirements in highly dynamic and complex propagation environments, antennas are being upgraded from traditional ones with fixed configurations and characteristics to reconfigurable ones with more flexible wave manipulation capabilities \cite{zhang2026survey,zhao2026reconfigurable,new2025}. 
By reconfiguring their electromagnetic characteristics, such as position, polarization, and radiation pattern, reconfigurable antennas can provide additional degrees of freedom to enhance system performance. 

Pixel antenna has emerged as one type of reconfigurable antennas that offers a promising solution for 6G from the physical layer \cite{pringle2004,chiu2012pixel}. Fundamentally, a pixel antenna discretizes a continuous radiating surface into an array of sub-wavelength metallic elements, termed pixels, which are interconnected via radio frequency (RF) switches. By electronically toggling the states of these switches, a process referred to as antenna coding in \cite{shen2024antenna}, the pixel antenna can be instantaneously reconfigured to adapt to the wireless channel. Such reconfigurability of the pixel antenna has been recently applied to exploit the spatial multiplexing \cite{han2024exploiting} so as to improve spectral efficiency for multi-input multi-output (MIMO) \cite{shen2024antenna} and multi-user systems \cite{li2025antenna}.  

Despite the profound advantages of pixel antennas, 
exploiting MIMO signal processing using pixel antennas is fundamentally different from and much more difficult than using conventional antennas with fixed configurations. 
This is due to the unique hardware constraints induced by pixel antennas that complicate not only the precoding design but, more importantly, also the channel state information (CSI) acquisition. 
Therefore, it is imperative and practical to design the pixel antenna empowered system under a limited feedback framework \cite{bejarano2013,Love2008limited}.
In such a paradigm, the receiver evaluates predefined codebooks of antenna coders using pixel antennas and digital precoders and feeds back low-overhead indices for data transmission, thereby eliminating the requirement of perfect CSI at the transmitter. 

In this work, we develop antenna coding and digital precoding solutions for limited feedback MIMO systems using pixel antennas, which leads to the following contributions. 
\emph{First}, we derive the optimal digital precoding that maximizes the spectral efficiency under practical power constraints. The derivation leads to a performance benchmark and, more importantly, provides insights on decoupling the joint antenna coding and precoding codebook construction in limited feedback scenarios. 
\emph{Second}, we propose a low-complexity offline codebook construction algorithm for both the antenna coder and the digital precoder to facilitate the limited feedback procedure. 
\emph{Third}, 
simulation results show that the proposed limited feedback systems using pixel antennas can significantly outperform unconstrained systems using fixed antennas and can approach unconstrained systems using pixel antennas with low quantization resolutions.

\emph{Notations:} $\mathbb{C}$ and $\mathbb{R}$ denote the set of complex and real numbers, respectively. Matrices, vectors, scalars, and sets are denoted by $\mathbf{A}$, $\mathbf{a}$, $a$, and $\mathcal{A}$, respectively. $(\cdot)^\mathsf{T}$, $(\cdot)^\mathsf{H}$, and $(\cdot)^{-1}$ denote the transpose, conjugate transpose, and inverse. $\|\cdot\|_2$ and $\|\cdot\|_\mathsf{F}$ are the $\ell_2$-norm and Frobenius norm, whereas $|\cdot|$ is the determinant. $[\mathbf{A}]_{:,1:N}$ and $[\mathbf{A}]_{i,i}$ are the first $N$ columns and the $i$th diagonal entry of $\mathbf{A}$, respectively. $\mathrm{diag}(\cdot)$ is a diagonal matrix with the given entries on its diagonal. $\mathrm{blkdiag}(\cdot)$ is a block diagonal matrix. $\mathbf{I}_N$ is the $N \times N$ identity matrix. $\mathbb{E}\{\cdot\}$ denotes expectation.

\section{Modeling of Pixel Antenna}
\begin{figure}[t]
    \centering
    \includegraphics[width=0.95\linewidth]{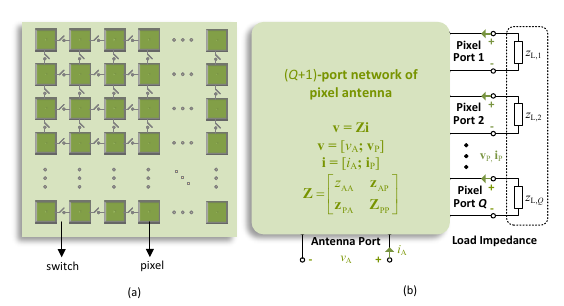} 
    \caption{(a) Schematic of pixel antenna. (b) Multiport circuit network model of pixel antenna.}
    \label{fig:antenna_model}
\end{figure}

The principle of pixel antennas involves the discretization of a continuous radiating surface into an array of smaller elements, named as pixels, and the interaction between pixels through RF switches \cite{pringle2004,chiu2012pixel}. As illustrated in Fig. \ref{fig:antenna_model}, RF switches between adjacent pixels enable dynamic connections, allowing for flexible adjustments to the antenna configuration and resulting a highly reconfigurable device.

Based on multiport network theory, a pixel antenna having $Q$ RF switches is modeled as a $(Q+1)$-port network, consisting of one antenna feed port (labeled ``A") and $Q$ pixel ports (labeled ``P") corresponding to the switches. The electromagnetic property of this network is characterized by an impedance matrix $\mathbf{Z} \in \mathbb{C}^{(Q+1)\times(Q+1)}$, partitioned as
$\mathbf{Z} = \begin{bmatrix} z_\mathrm{AA} & \mathbf{z}_\mathrm{AP} \\ \mathbf{z}_\mathrm{PA} & \mathbf{Z}_\mathrm{PP} \end{bmatrix}$,
where $z_\mathrm{AA}$ is the self-impedance of the feed port, $\mathbf{Z}_\mathrm{PP} \in \mathbb{C}^{Q \times Q}$ is the impedance matrix of the pixel ports, $\mathbf{z}_\mathrm{PA} \in \mathbb{C}^{Q \times 1}$ represents the trans-impedance between the feed and pixel ports, and $\mathbf{z}_\mathrm{AP} = \mathbf{z}_\mathrm{PA}^\mathsf{T}$. The impedance matrix relates the voltages and currents at all the $(Q+1)$ ports by $\mathbf{v} = \mathbf{Z}\mathbf{i}$, where $\mathbf{v} = [v_\mathrm{A};\mathbf{v}_\mathrm{P}] $ concatenates the voltages $v_\mathrm{A}\in\mathbb{C}$ and $\mathbf{v}_\mathrm{P}\in\mathbb{C}^{Q\times 1}$ respectively at antenna and pixel ports, and  $\mathbf{i} = [i_\mathrm{A};\mathbf{i}_\mathrm{P}] $ concatenates the currents $i_\mathrm{A}\in\mathbb{C}$ and $\mathbf{i}_\mathrm{P}\in\mathbb{C}^{Q\times 1}$ respectively at antenna and pixel ports.

As illustrated in Fig. \ref{fig:antenna_model}, $Q$ pixel ports are further one-to-one connected to $Q$ load impedances $z_{\mathrm{L},1},\ldots,z_{\mathrm{L},Q}$. These load impedances are either short-circuited, i.e. $z_{\mathrm{L},q} = 0$, or open-circuited, i.e.  $z_{\mathrm{L},q} = \infty$, to indicate the on and off states of each RF switch. The states of these $Q$ switches can be conveniently mapped to a binary vector $\mathbf{b} = [b_1,\ldots,b_Q]^\mathsf{T}\in\{0,1\}^{Q\times 1}$, named the antenna coder, with $z_{\mathrm{L},q} = 0$ if $b_q = 0$ and  $z_{\mathrm{L},q} = \infty$ if $b_q = 1$.
Define the load impedance matrix as $\mathbf{Z}_\mathrm{L}(\mathbf{b}) = \mathrm{diag}(z_{\mathrm{L},1},\ldots,z_{\mathrm{L},Q})$ which relates the voltage $\mathbf{v}_\mathrm{P}$ and current $\mathbf{i}_\mathrm{P}$ at pixel ports by $\mathbf{v}_\mathrm{P} = -\mathbf{Z}_{\mathrm{L}}(\mathbf{b})\mathbf{i}_\mathrm{P}$. Then $\mathbf{i}_\mathrm{P}$ is strictly coupled to the current $i_\mathrm{A}$ via the following constraint
\begin{equation} \label{eq:current_coupling}
    \mathbf{i}_\mathrm{P}(\mathbf{b}) = -(\mathbf{Z}_\mathrm{PP} + \mathbf{Z}_\mathrm{L}(\mathbf{b}))^{-1} \mathbf{z}_\mathrm{PA} i_\mathrm{A}.
\end{equation}

With the coded currents $\mathbf{i}(\mathbf{b}) = [i_\mathrm{A};\mathbf{i}_\mathrm{P}(\mathbf{b})]$,  we obtain the corresponding coded radiation pattern of the pixel antenna, which is formed by superposing the radiation patterns from all $Q+1$ ports weighted by $\mathbf{i}(\mathbf{b})$ as
\begin{equation}
    \mathbf{e}(\mathbf{b}) = \mathbf{E}_\mathrm{oc} \, \mathbf{i}(\mathbf{b}),
\end{equation}
where $\mathbf{E}_\mathrm{oc} = [\mathbf{e}_\mathrm{A}^\mathrm{oc}, \mathbf{e}_{\mathrm{P},1}^\mathrm{oc}, \dots, \mathbf{e}_{\mathrm{P},Q}^\mathrm{oc}] \in \mathbb{C}^{2K \times (Q+1)}$ denotes the open-circuit radiation pattern matrix. This matrix collects the individual radiation patterns $\mathbf{e}_{\mathrm{A}}^\mathrm{oc} \in \mathbb{C}^{2K\times 1}$ for the antenna port and $\mathbf{e}_{\mathrm{P},q}^\mathrm{oc} \in \mathbb{C}^{2K\times 1}$ for the $q$th pixel port, $\forall q = 1, \ldots, Q$ including both $\theta$- and $\phi$-polarization components over $K$ spatial samples, excited by a unit current at that port while all other ports are left open-circuited. Consequently, by flexibly selecting the antenna coder $\mathbf{b}$ within its feasible range, the radiation pattern $\mathbf{e}(\mathbf{b})$ of the pixel antenna can be effectively reconfigured, thereby offering additional beam control flexibility for wireless systems.

\section{System Model}
We consider a downlink MIMO communication system operating in a rich-scattered environment. The transmitter is equipped with $N_\mathrm{T}$ pixel antennas, each of which is coded by an antenna coder $\mathbf{b}_n$, $\forall n = 1,\ldots,N_\mathrm{T}$, while the receiver is equipped with $N_\mathrm{R}$ conventional antennas with fixed configurations. The system transmits $N_\mathrm{S}$ data streams, satisfying $N_\mathrm{S} \le \min\{N_\mathrm{T},N_\mathrm{R}\}$. In this study, we focus on the scenario where $N_\mathrm{T} = N_\mathrm{R} = N$, and thus $N_\mathrm{S} \le N$.
The data stream vector $\mathbf{s} = [s_1,\ldots,s_{N_\mathrm{S}}]^\mathsf{T}\in\mathbb{C}^{N_\mathrm{S}\times 1}$ satisfying $\mathbb{E}\{\mathbf{s}\mathbf{s}^\mathsf{H}\} = \frac{P}{N_\mathrm{S}}\mathbf{I}_{N_\mathrm{S}}$ with $P$ being the transmit power is precoded by a digital precoder matrix $\mathbf{F}_\mathrm{D}\in\mathbb{C}^{N\times N_\mathrm{S}}$ subject to a normalized power constraint $\|\mathbf{F}_\mathrm{D}\|_\mathsf{F}^2\le N_\mathrm{S}$. Hence, the baseband transmit signal is given by $\mathbf{x} = \mathbf{F}_\mathrm{D}\mathbf{s}$.

\subsection{Channel Model}
The beamspace channel representation of the considered MIMO system is given by \cite{pillai2012array}
\begin{equation}
    \mathbf{H}(\mathbf{B}) = \mathbf{E}_\mathrm{R}^\mathsf{T} \mathbf{H}_\mathrm{V} \mathbf{E}_\mathrm{T}(\mathbf{B}),
\end{equation}
where $\mathbf{B} = [\mathbf{b}_1, \dots, \mathbf{b}_{N}]$ represents the antenna coder matrix at the transmitter, $\mathbf{E}_\mathrm{R}\in\mathbb{C}^{2K\times N}$ is the receiver radiation pattern matrix satisfying $\mathbf{E}_\mathrm{R}^\mathsf{H}\mathbf{E}_\mathrm{R} = \mathbf{I}_{N}$ based on the assumption that receive antennas are spatially uncorrelated and have orthogonal radiation patterns. 
$\mathbf{H}_\mathrm{V}\in\mathbb{C}^{2K\times 2K}$ is the virtual propagation channel with each entry being the channel gain from an angle of departure to an angle of arrival across $K$ spatial angle  samples for $\theta$ and $\phi$ polarizations, and $\mathbf{E}_\mathrm{T}(\mathbf{B}) = [\mathbf{e}_1(\mathbf{b}_1),\ldots,\mathbf{e}_n(\mathbf{b}_{N})]\in\mathbb{C}^{2K\times N}$ denotes the transmitter radiation pattern matrix with the $n$th column $\mathbf{e}_n(\mathbf{b}_n) = \mathbf{E}_{\mathrm{oc},n}\mathbf{i}_n(\mathbf{b}_n)$ being the normalized radiation pattern of the $n$th pixel antenna at the transmitter, i.e. $\|\mathbf{e}(\mathbf{b}_n)\|_2^2 = 1$, $\forall n = 1, \ldots, N$.

The effective rank\footnote{We assume all pixel antennas are identical so that they have the same effective aerial degrees of freedom.} of $\mathbf{E}_{\mathrm{oc},n}$, denoted as $N_\mathrm{eff}$, is known as the effective aerial degrees of freedom \cite{Han2021effective} that identifies the number of orthogonal radiation patterns that the antenna can provide. This can be mathematically illustrated by performing the singular value decomposition (SVD) on the open-circuit pattern matrix $\mathbf{E}_{\mathrm{oc},n}$ as
\begin{equation}
    \mathbf{E}_{\mathrm{oc},n} = \mathbf{U}_n\mathbf{S}_n\mathbf{V}_n^\mathsf{H},
\end{equation}
where $\mathbf{U}_n\in\mathbb{C}^{2K\times N_\mathrm{eff}}$ and $\mathbf{V}_n\in\mathbb{C}^{(Q+1)\times N_\mathrm{eff}}$ are semi-unitary matrices and $\mathbf{S}_n\in\mathbb{C}^{N_\mathrm{eff}\times N_\mathrm{eff}}$ collects dominant singular values in the diagonal. This allows us to rewrite the MIMO channel as the following form
\begin{equation}
    \mathbf{H}(\mathbf{B}) = \mathbf{H}_\mathrm{eff}\mathbf{F}(\mathbf{B}),\label{eq:channel}
\end{equation}
where $\mathbf{H}_\mathrm{eff} = [\mathbf{H}_{\mathrm{eff},1},\ldots,\mathbf{H}_{\mathrm{eff},N}]\in\mathbb{C}^{N\times NN_\mathrm{eff}}$ with each block being $\mathbf{H}_{\mathrm{eff},n} = \mathbf{E}_\mathrm{R}^\mathsf{T} \mathbf{H}_\mathrm{V} \mathbf{U}_n$ and 
$\mathbf{F}(\mathbf{B}) = \mathrm{blkdiag}(\mathbf{f}_1(\mathbf{b}_1),\ldots,\mathbf{f}_N(\mathbf{b}_N))\in\mathbb{C}^{NN_\mathrm{eff}\times N}$ with each block being $\mathbf{f}_n(\mathbf{b}_n) = \mathbf{S}_n \mathbf{V}_n^\mathsf{H}\mathbf{i}_n(\mathbf{b}_n)$. 
From equation (\ref{eq:channel}) we observe that $\mathbf{H}_\mathrm{eff}$ acts as an effective fixed MIMO channel, while $\mathbf{F}(\mathbf{B})$ serves as an analog pattern coder controlled by the antenna coder matrix $\mathbf{B}$.

\subsection{Problem Formulation}

We consider a downlink MIMO system where the transmitter and the receiver are connected via a limited feedback link. Here we assume that the receiver has perfect CSI and selects the best antenna coder and digital precoder matrices $\mathbf{B}^\star$ and $\mathbf{F}_\mathrm{D}^\star$ from pre-defined codebooks aiming at maximizing the spectral efficiency. Based on this setup, we first write the received signal vector as
\begin{equation}
    \mathbf{y} = \mathbf{H}(\mathbf{B})\mathbf{x} + \mathbf{n} = \mathbf{H}_\mathrm{eff}\mathbf{F}(\mathbf{B})\mathbf{F}_\mathrm{D}\mathbf{s} + \mathbf{n},
\end{equation}
where $\mathbf{n} \sim \mathcal{CN}(\mathbf{0}, \sigma^2 \mathbf{I}_{N})$ is the noise. The spectral efficiency of the system is
\begin{equation} \label{eq:mutual_info}
    \mathcal{I}(\mathbf{B}, \mathbf{F}_\mathrm{D}) = \log_2\left|\mathbf{I}_{N} + \frac{P}{N_\mathrm{S}\sigma^2} \mathbf{H}_\mathrm{eff} \mathbf{F}(\mathbf{B}) \mathbf{F}_\mathrm{D} \mathbf{F}_\mathrm{D}^\mathsf{H} \mathbf{F}^\mathsf{H}(\mathbf{B}) \mathbf{H}_\mathrm{eff}^\mathsf{H}\right|.
\end{equation}

In the limited feedback scenario, the antenna coder matrix $\mathbf{B}$ and the digital precoder matrix $\mathbf{F}_\mathrm{D}$ are respectively taken from codebooks $\mathcal{B}$ and $\mathcal{F}_\mathrm{D}$. In this sense, the spectral efficiency maximization problem is formulated as
\begin{subequations} \label{eq:opt_problem}
\begin{align}
    \max \quad & \mathcal{I}(\mathbf{B}, \mathbf{F}_\mathrm{D})\\
    \text{s.t.} \quad & \mathbf{B} \in \mathcal{B},\\
    & \mathbf{F}_\mathrm{D} \in \mathcal{F}_\mathrm{D}.
\end{align}
\end{subequations}

The objective in this work is to construct efficient and effective codebooks $\mathcal{B}$ and $\mathcal{F}_\mathrm{D}$ to maximize the spectral efficiency in (8). To facilitate the codebook construction, we will start from the case where only the antenna coder matrix is selected from the codebook while the digital precoder can be optimally determined. This will provide insights on simplifying the joint codebook construction problem, as will be detailed in the following section.

\section{Optimal Digital Precoder Design}
\label{sec:opt}
We first perform SVD on the effective channel $\mathbf{H}_\mathrm{eff}$ as 
\begin{equation}
    \mathbf{H}_\mathrm{eff} = \bar{\mathbf{U}} \bar{\mathbf{S}} \bar{\mathbf{V}}^\mathsf{H}, 
\end{equation}
where $\bar{\mathbf{U}} \in \mathbb{C}^{N \times N}$ and $\bar{\mathbf{V}} \in \mathbb{C}^{NN_\mathrm{eff} \times N}$ are semi-unitary matrices, and $\bar{\mathbf{S}} \in \mathbb{R}^{N \times N}$ is a diagonal matrix containing the singular values. 
We then extract the core part of the channel $\mathbf{H}(\mathbf{B})$, and perform SVD on it as 
\begin{equation}
    \bar{\mathbf{S}}\bar{\mathbf{V}}^\mathsf{H}\mathbf{F}(\mathbf{B}) = \bar{\bar{\mathbf{U}}} \bar{\bar{\mathbf{S}}} \bar{\bar{\mathbf{V}}}^\mathsf{H},
\end{equation} 
where $\bar{\bar{\mathbf{U}}} \in \mathbb{C}^{N\times N}$ and $\bar{\bar{\mathbf{V}}} \in \mathbb{C}^{N \times N}$ are unitary matrices, and $\bar{\bar{\mathbf{S}}} \in \mathbb{R}^{N\times N}$ contains the singular values of the channel $\mathbf{H}(\mathbf{B})$. Based on this exact decomposition, we investigate the optimal digital precoder under the following two different practical power constraints.

\subsection{Total Power Constraint}
\label{subsec:opt_tot}
Under the total transmit power constraint $\|\mathbf{F}_\mathrm{D}\|_\mathsf{F}^2 \le N_\mathrm{S}$, it is a well-established result that the optimal digital precoder aligns with the dominant right singular vectors of the channel. Therefore, the optimal digital precoder is given by
\begin{equation} \label{eq:opt_FD}
    \mathbf{F}_\mathrm{D}^\star = [\bar{\bar{\mathbf{V}}}]_{:, 1:N_\mathrm{S}} \mathbf{\Lambda},
\end{equation}
where $[\bar{\bar{\mathbf{V}}}]_{:, 1:N_\mathrm{S}} \in \mathbb{C}^{N \times N_\mathrm{S}}$ collects the first $N_\mathrm{S}$ singular vectors corresponding to $N_\mathrm{S}$ dominant singular values and $\mathbf{\Lambda} = \mathrm{diag}(\sqrt{p_1}, \dots, \sqrt{p_{N_\mathrm{S}}}) \in \mathbb{C}^{N_\mathrm{S} \times N_\mathrm{S}}$ with $p_i\ge 0$ is the diagonal power allocation matrix. 

The allocated power $p_i$ for the $i$th data stream is determined by the standard water-filling algorithm as
\begin{equation}
    p_i = \left( \mu - \frac{N_\mathrm{S}\sigma^2}{P\bar{\bar{\sigma}}_i^2} \right)^+, \quad \forall i = 1,\ldots,N_\mathrm{S},
\end{equation}
where $\bar{\bar{\sigma}}_i = [\bar{\bar{\mathbf{S}}}]_{i,i}$ denotes the $i$th dominant singular value of the channel, and $(a)^+ = \max(0, a)$. The Lagrangian multiplier $\mu$ is strictly chosen to satisfy the total transmit power constraint
$\sum_{i=1}^{N_\mathrm{S}} ( \mu - \frac{N_\mathrm{S}\sigma^2}{P\bar{\bar{\sigma}}_i^2} )^+ = N_\mathrm{S}$.

Substituting the optimal $\mathbf{F}_\mathrm{D}^\star$ back into the mutual information expression in (\ref{eq:mutual_info}), the maximum achievable mutual information for a given antenna coder $\mathbf{B}$ is formulated as
\begin{equation}\label{eq:optimal_I}
    \mathcal{I}^\star(\mathbf{B}) = \log_2 \left| \mathbf{I}_{N_\mathrm{S}} + \frac{P}{N_\mathrm{S}\sigma^2} [\bar{\bar{\mathbf{S}}}]_{1:N_\mathrm{S}, 1:N_\mathrm{S}}^2 \mathbf{\Lambda}^2 \right|.
\end{equation}
Since both the power allocation $p_i$ and the singular values $\bar{\bar{\sigma}}_i$ are entirely determined by $\mathbf{F}(\mathbf{B})$, the optimal spectral efficiency is solely a function of the selected antenna coder. In this sense, assuming the availability of a continuous digital precoder, the absolute performance upper bound can be simply obtained by exhaustively searching the codebook $\mathcal{B}$ and finding the index that maximizes (\ref{eq:optimal_I}). 

\subsection{Unitary Power Constraint}
While the total power constraint addressed by water-filling provides the theoretical capacity upper bound, it imposes significant complexity on the codebook design in limited feedback scenarios, as both the spatial directions and the continuous power allocations in the digital precoder must be quantized. As an alternative, a unitary power constraint is widely considered in practical limited feedback MIMO systems \cite{Love2008limited}. This constraint enforces the digital precoder to have orthogonal columns with uniform power allocation. 
Even though possible performance loss is expected compared to the relaxed total power constraint, digital precoders with unitary constraints usually lead to more efficient codebooks and codeword selection algorithms \cite{Alk2016hybrid}. This will be proven in the following derivations and performance evaluations.

In this case, equal power is allocated across all active streams, yielding $p_1 = \ldots = p_{N_\mathrm{S}} = 1$. The optimal digital precoder simplifies to $\mathbf{F}_\mathrm{D}^\star = [\bar{\bar{\mathbf{V}}}]_{:, 1:N_\mathrm{S}}$, and the corresponding mutual information for a given $\mathbf{B}$ becomes 
\begin{equation} \label{eq:optimal_I_unitary}
    \mathcal{I}_\mathrm{uni}^\star(\mathbf{B}) = \log_2 \left| \mathbf{I}_{N_\mathrm{S}} + \frac{P}{N_\mathrm{S}\sigma^2} [\bar{\bar{\mathbf{S}}}]_{1:N_\mathrm{S}, 1:N_\mathrm{S}}^2 \right|.
\end{equation}
This simplified metric acts as the mathematical foundation for our tractable joint codebook design, which will be detailed in the subsequent section.

\section{Codebook Design for Antenna Coder and Digital Precoder}
\label{sec:codebook}
In this section, we design the codebook for the proposed limited feedback MIMO system. We  consider two cases based on the relationship between $N$ and $N_\mathrm{S}$ and respectively construct codebooks for antenna coder and digital precoder.

\subsection{Case 1: $N_\mathrm{S} = N$}
For the special case of $N_\mathrm{S} = N$, the digital precoder $\mathbf{F}_\mathrm{D}$ becomes an orthogonal matrix satisfying $\mathbf{F}_\mathrm{D}\mathbf{F}_\mathrm{D}^\mathsf{H} = \mathbf{I}_{N}$. Therefore, the spectral efficiency optimization objective (\ref{eq:mutual_info}) can be simplified as
\begin{align}\label{eq:simplified_I}
    \mathcal{I}_\mathrm{uni}(\mathbf{B},\mathbf{F}_\mathrm{D}) &= 
    \log_2 \left| \mathbf{I}_{N} + \frac{P}{N_\mathrm{S}\sigma^2} \bar{\mathbf{U}}\bar{\mathbf{S}} \bar{\mathbf{V}}^\mathsf{H} \mathbf{F}(\mathbf{B})\right. \nonumber \\
    &\qquad \qquad \qquad\left.\times \mathbf{F}_\mathrm{D}\mathbf{F}_\mathrm{D}^\mathsf{H} \mathbf{F}^\mathsf{H}(\mathbf{B}) \bar{\mathbf{V}} \bar{\mathbf{S}} \bar{\mathbf{U}}^\mathsf{H} \right| \nonumber \\
    &= \log_2 \left| \mathbf{I}_{N} + \frac{P}{N_\mathrm{S} \sigma^2} \bar{\mathbf{S}} \bar{\mathbf{V}}^\mathsf{H} \mathbf{F}(\mathbf{B}) \mathbf{F}^\mathsf{H}(\mathbf{B}) \bar{\mathbf{V}}\bar{\mathbf{S}} \right| \nonumber\\
    &= \log_2\left|\mathbf{I}_N + \frac{P}{N_\mathrm{S}\sigma^2}\bar{\bar{\mathbf{S}}}^2\right|
    = \mathcal{I}^\star_\mathrm{uni}(\mathbf{B}). 
\end{align}
This simplification reveals that the maximum spectral efficiency can be obtained by selecting solely the antenna coder matrix $\mathbf{B}$ that maximizes the projection of radiation patterns of pixel antennas onto the dominant subspaces of the effective channel $\mathbf{H}_\mathrm{eff}$.
In other words, in the special case of $N = N_\mathrm{S}$, the optimal limited feedback spectral efficiency is independent of the value of the digital precoder, indicating that there is no need to quantize $\mathbf{F}_\mathrm{D}$. Therefore, we focus on the construction of the antenna coder codebook $\mathcal{B}$ and propose a heuristic algorithm based on the generalized Lloyd method \cite{Gersho1992Vector} and the successive exhaustive Boolean optimization \cite{Shen2016Successive}. The detailed procedure is summarized in Algorithm \ref{alg:codebook}.

\begin{algorithm}[t]
\caption{Antenna Coder Codebook Construction}
\label{alg:codebook}
\begin{enumerate}
\item\textbf{Initialization:} Generate an initial codebook with $M$ random codewords as $\mathcal{B} = \{\mathbf{B}_1, \dots, \mathbf{B}_M\}$ and $S$ training channel samples $\mathcal{H}_\mathrm{train} = \{\mathbf{H}_\mathrm{V}^{s}\}_{s=1}^S$.
\item\textbf{Data Processing:} Calculate effective channel samples $\{\mathbf{H}_\mathrm{eff}^{s}\}_{s=1}^S$ and construct the set $\mathcal{V}_\mathrm{train} =\{(\bar{\mathbf{S}}^{s},\bar{\mathbf{V}}^{s})\}_{s=1}^S$ by performing SVD $\mathbf{H}_\mathrm{eff}^s = \bar{\mathbf{U}}^s\bar{\mathbf{S}}^s(\bar{\mathbf{V}}^s)^\mathsf{H}$. 
\item\textbf{Nearest Neighbor Partitioning:} 
Partition the training set $\mathcal{V}_\mathrm{train}$ into $M$ Voronoi regions $\mathcal{R}_1, \dots, \mathcal{R}_M$. Specifically, $\mathcal{R}_m$ refers to the nearest neighbor of codeword $\mathbf{B}_m$ as 
\begin{equation}
    \begin{split}
        \mathcal{R}_m = &\left\{ (\bar{\mathbf{S}}^{s},\bar{\mathbf{V}}^{s}) \;\middle|\; \mathcal{I}_\mathrm{uni}(\bar{\mathbf{S}}^{s},\bar{\mathbf{V}}^{s},\mathbf{B}_m) \right. \\
        &~~~~\left. \ge \mathcal{I}_\mathrm{uni}(\bar{\mathbf{S}}^{s},\bar{\mathbf{V}}^{s},\mathbf{B}_{m'}), \forall m \ne m', \forall s \right\},
    \end{split}
\end{equation}
where $\mathcal{I}_\mathrm{uni}(\bar{\mathbf{S}}^{s},\bar{\mathbf{V}}^{s},\mathbf{B}_m)$ is modified from (\ref{eq:simplified_I}) as
\begin{equation} \label{eq:I_unitary_antenna_coder}
    \begin{aligned}
        &\mathcal{I}_\mathrm{uni}(\bar{\mathbf{S}}^{s},\bar{\mathbf{V}}^{s},\mathbf{B}_m)\\
         &~~~= \log_2 \left| \mathbf{I}_{N_\mathrm{S}} + \rho (\bar{\mathbf{S}}^s)^2(\bar{\mathbf{V}}^s)^\mathsf{H}\mathbf{F}(\mathbf{B}_m)\mathbf{F}^\mathsf{H}(\mathbf{B}_m)\bar{\mathbf{V}}^s\right|.
    \end{aligned}
\end{equation}
with $\rho$ a constant to generate a codebook commonly used for different transmit power budgets.
\item\textbf{Centroid Update (Codeword Optimization):}
For each region $\mathcal{R}_m$, update the codeword $\mathbf{B}_m$ to maximize the average spectral efficiency as
\begin{equation} \label{eq:centroid_obj}
    \mathbf{B}_m = \arg \max_{\mathbf{B} \in \{0,1\}^{Q \times N}} \sum_{(\bar{\mathbf{S}}^{s},\bar{\mathbf{V}}^{s}) \in \mathcal{R}_m} \mathcal{I}_\mathrm{uni}(\bar{\mathbf{S}}^{s},\bar{\mathbf{V}}^{s},\mathbf{B}).
\end{equation}
Problem (\ref{eq:centroid_obj}) is a binary optimization that can be directly solved by the SEBO \cite{Shen2016Successive}. 
\item\textbf{Loop back to step 3)} until convergence.
\end{enumerate}
\end{algorithm}

\subsection{Case 2: $N_\mathrm{S} < N$}

In the scenario where the number of data streams is less than the number of transmit antennas, the digital precoder $\mathbf{F}_\mathrm{D} \in \mathbb{C}^{N \times N_\mathrm{S}}$ is not a square matrix, meaning $\mathbf{F}_\mathrm{D} \mathbf{F}_\mathrm{D}^\mathsf{H} \neq \mathbf{I}_{N}$.
Consequently, the analog pattern coder $\mathbf{F}(\mathbf{B})$ and the digital precoder $\mathbf{F}_\mathrm{D}$ are coupled and thus both $\mathbf{B}$ and $\mathbf{F}_\mathrm{D}$ should be quantized and fed back to the transmitter.
To facilitate a tractable offline codebook design, we propose to decouple the impact of antenna coder $\mathbf{B}$ and digital precoder $\mathbf{F}_\mathrm{D}$ and thus simplify the joint codebook construction problem. More details will be provided in the sequel.

\subsubsection{Construction of Antenna Coder Codebook $\mathcal{B}$}
To derive a performance metric as a function of only $\mathbf{B}$, we exploit the optimal digital precoder design in Section III-B.
Assuming the digital precoder is optimized without quantization, we have the following spectral efficiency from (\ref{eq:optimal_I_unitary}) that can be further approximated as
\begin{align}
    \mathcal{I}_\mathrm{uni}^\star(\mathbf{B}) &= \log_2 \left| \mathbf{I}_{N_\mathrm{S}} + \frac{P}{N_\mathrm{S} \sigma^2} [\bar{\bar{\mathbf{S}}}]_{1:N_\mathrm{S}, 1:N_\mathrm{S}}^2 \right| \nonumber \\
    &= \sum_{i=1}^{N_\mathrm{S}} \log_2 \left( 1 + \frac{P}{N_\mathrm{S} \sigma^2} \lambda_i \left( \bar{\mathbf{S}} \bar{\mathbf{V}}^\mathsf{H} \mathbf{F}(\mathbf{B}) \mathbf{F}^\mathsf{H}(\mathbf{B}) \bar{\mathbf{V}} \bar{\mathbf{S}} \right) \right) \nonumber \\
    &\ge \sum_{i=1}^{N_\mathrm{S}} \log_2 \left( 1 + \frac{P}{N_\mathrm{S} \sigma^2} \lambda_i \left( \tilde{\mathbf{S}} \tilde{\mathbf{V}}^\mathsf{H} \mathbf{F}(\mathbf{B}) \mathbf{F}^\mathsf{H}(\mathbf{B}) \tilde{\mathbf{V}} \tilde{\mathbf{S}} \right) \right) \nonumber \\
    &= \log_2 \left| \mathbf{I}_{N_\mathrm{S}} + \frac{P}{N_\mathrm{S} \sigma^2} \tilde{\mathbf{S}} \tilde{\mathbf{V}}^\mathsf{H} \mathbf{F}(\mathbf{B}) \mathbf{F}^\mathsf{H}(\mathbf{B}) \tilde{\mathbf{V}} \tilde{\mathbf{S}} \right| \nonumber \\
    &= \tilde{\mathcal{I}}_\mathrm{uni}(\mathbf{B}), \label{eq:I_uni_18}
\end{align}
where $\tilde{\mathbf{S}} = [\bar{\mathbf{S}}]_{1:N_\mathrm{S}, 1:N_\mathrm{S}} \in \mathbb{R}^{N_\mathrm{S} \times N_\mathrm{S}}$ denotes the truncated diagonal singular value matrix capturing the primary $N_\mathrm{S}$ modes of the effective channel, $\tilde{\mathbf{V}} = [\bar{\mathbf{V}}]_{:, 1:N_\mathrm{S}} \in \mathbb{C}^{N_\mathrm{eff} \times N_\mathrm{S}}$ represents the corresponding right-singular vectors, and $\lambda_i(\cdot)$ denotes the $i$th eigenvalue function. 

Now we establish a metric as a function of only $\mathbf{B}$ that allows us to construct the antenna coder codebook by maximizing the performance lower bound $\tilde{\mathcal{I}}_\mathrm{uni}(\mathbf{B})$.
With this newly derived metric, the codebook construction procedure proposed in Algorithm \ref{alg:codebook} can be seamlessly adapted for the case of $N_\mathrm{S} < N$.
This can be simply done by substituting the objective $\mathcal{I}_\mathrm{uni}(\bar{\mathbf{S}}^{s},\bar{\mathbf{V}}^{s},\mathbf{B}_m)$ in (\ref{eq:I_unitary_antenna_coder}) with 
\begin{equation}
    \begin{aligned}
    &\tilde{\mathcal{I}}_\mathrm{uni}(\tilde{\mathbf{S}}^{s},\tilde{\mathbf{V}}^{s},\mathbf{B}_m)\\
     &~~~= \log_2 \left| \mathbf{I}_{N_\mathrm{S}} + \rho \tilde{\mathbf{S}}^s (\tilde{\mathbf{V}}^s)^\mathsf{H} \mathbf{F}(\mathbf{B}_m) \mathbf{F}^\mathsf{H}(\mathbf{B}_m) \tilde{\mathbf{V}}^s \tilde{\mathbf{S}}^s \right|,
    \end{aligned}
\end{equation} 
in both the nearest neighbor partitioning and the SEBO-based centroid update of Algorithm \ref{alg:codebook}. 

\subsubsection{Construction of Digital Precoder Codebook $\mathcal{F}_\mathrm{D}$}
When the optimal antenna coder $\mathbf{B}^\star$ is chosen from the codebook $\mathcal{B}$, the MIMO channel $\mathbf{H}(\mathbf{B}^\star) = \mathbf{H}_\mathrm{eff}\mathbf{F}(\mathbf{B}^\star)$ is stabilized. We then focus on the following conditional metric to construct the digital precoder codebook as 
\begin{equation} \label{eq:cond_FD}
    \mathcal{I}_\mathrm{uni}(\mathbf{F}_\mathrm{D}~|~\mathbf{B}^\star) = \log_2 \left| \mathbf{I}_{N} + \frac{P}{N_\mathrm{S}\sigma^2}\mathbf{H}(\mathbf{B}^\star)\mathbf{F}_\mathrm{D}\mathbf{F}_\mathrm{D}^\mathsf{H}\mathbf{H}^\mathsf{H}(\mathbf{B}^\star) \right|. 
\end{equation}

Now we establish a metric as a function of only $\mathbf{F}_\mathrm{D}$ under the condition that the antenna coder matrix has been determined based on the codebook $\mathcal{B}$. 
To generate the digital precoder codebook $\mathcal{F}_\mathrm{D}$, we again use the generalized Lloyd method and propose a heuristic codebook construction algorithm as summarized in Algorithm \ref{alg:codebook_dig}.

\begin{algorithm}[t]
\caption{Digital Precoder Codebook Construction}
\label{alg:codebook_dig}
\begin{enumerate}
\item\textbf{Initialization:} Generate an initial codebook with $M$ random codewords as $\mathcal{F}_\mathrm{D} = \{\mathbf{F}_{\mathrm{D},1}, \dots, \mathbf{F}_{\mathrm{D},M}\}$ and $S$ training channel samples $\mathcal{H}_\mathrm{train} = \{\mathbf{H}_\mathrm{V}^{s}\}_{s=1}^S$.
\item\textbf{Data Processing:} Calculate effective channel samples $\mathcal{H}_\mathrm{eff} = \{\mathbf{H}_\mathrm{eff}^{s}\}_{s=1}^S$ and perform $\mathbf{H}_\mathrm{eff}^s = \bar{\mathbf{U}}^s\bar{\mathbf{S}}^s(\bar{\mathbf{V}}^s)^\mathsf{H}$. Obtain $\tilde{\mathbf{S}}^s = [\bar{\mathbf{S}}]_{1:N_\mathrm{S},1:N_\mathrm{S}}$, $\tilde{\mathbf{V}}^s = [\bar{\mathbf{V}}^s]_{:,1:N_\mathrm{S}}$. 
\item \textbf{Antenna Coder Design:} Obtain antenna coders for effective channel samples by solving 
\begin{equation}
    \mathbf{B}^s = \arg~\max_{\mathbf{B}\in\mathcal{B}}~\tilde{\mathcal{I}}_\mathrm{uni}(\tilde{\mathbf{S}}^s,\tilde{\mathbf{V}}^s,\mathbf{B}),
\end{equation}
with an exhaustive search over $\mathcal{B}$ from Algorithm \ref{alg:codebook}.
\item\textbf{Nearest Neighbor Partitioning:} 
Partition the training set $\mathcal{H}_\mathrm{eff}$ into $M$ Voronoi regions $\mathcal{R}_1, \dots, \mathcal{R}_M$. Specifically, $\mathcal{R}_m$ refers to the nearest neighbor of codeword $\mathbf{F}_{\mathrm{D},m}$ as 
\begin{equation}
    \begin{split}
        \mathcal{R}_m = &\left\{ \mathbf{H}_\mathrm{eff}^s \;\middle|\; \mathcal{I}_\mathrm{uni}(\mathbf{H}_\mathrm{eff}^s,\mathbf{F}_{\mathrm{D},m}~|~\mathbf{B}^s) \right. \\
        &\left. \ge \mathcal{I}_\mathrm{uni}(\mathbf{H}_\mathrm{eff}^s,\mathbf{F}_{\mathrm{D},m'}~|~\mathbf{B}^s), \forall m \ne m', \forall s \right\},
    \end{split}
\end{equation}
where $\mathcal{I}_\mathrm{uni}(\mathbf{H}_\mathrm{eff}^s,\mathbf{F}_{\mathrm{D},m}~|~\mathbf{B}^s)$ is modified from (\ref{eq:cond_FD}) as
\begin{equation} 
    \begin{aligned}
        &\mathcal{I}_\mathrm{uni}(\mathbf{H}_\mathrm{eff}^s,\mathbf{F}_{\mathrm{D},m}~|~\mathbf{B}^s)\\ 
        &~ = \log_2\left|\mathbf{I}_{N} + \rho \mathbf{H}_\mathrm{eff}^s \mathbf{F}(\mathbf{B}^s) \mathbf{F}_{\mathrm{D},m} \mathbf{F}_{\mathrm{D},m}^\mathsf{H} \mathbf{F}^\mathsf{H}(\mathbf{B}^s) (\mathbf{H}_\mathrm{eff}^s)^\mathsf{H}\right|.
    \end{aligned}
\end{equation}
\item\textbf{Centroid Update (Codeword Optimization):}
For each region $\mathcal{R}_m$, update the codeword $\mathbf{F}_{\mathrm{D},m}$ to maximize the average spectral efficiency as
\begin{equation} 
    \mathbf{F}_{\mathrm{D},m} = \arg\max_{\mathbf{F}_\mathrm{D}^\mathsf{H}\mathbf{F}_\mathrm{D} = \mathbf{I}_{N_\mathrm{S}}} \sum_{\mathbf{H}_{\mathrm{eff}}^s\in\mathcal{R}_m}\mathcal{I}_\mathrm{uni}(\mathbf{H}_\mathrm{eff}^s,\mathbf{F}_{\mathrm{D}}~|~\mathbf{B}^s),
\end{equation}
which is approximated into an average chordal distance minimization problem \cite{Alk2016hybrid} between $\mathbf{F}_{\mathrm{D},m}$ and optimal digital precoders $\mathbf{F}_\mathrm{D}^s\in\mathbb{C}^{N\times N_\mathrm{S}}$ for each effective channel samples in $\mathcal{R}_m$ as 
\begin{equation}
    \begin{aligned}
        \mathbf{F}_{\mathrm{D},m} &= \arg\min_{\mathbf{F}_\mathrm{D}^\mathsf{H}\mathbf{F}_\mathrm{D} = \mathbf{I}_{N_\mathrm{S}}} \sum_{\mathbf{H}_{\mathrm{eff}}^s\in\mathcal{R}_m} (N_\mathrm{S} - \|\mathbf{F}_\mathrm{D}^\mathsf{H}\mathbf{F}_\mathrm{D}^s\|_\mathsf{F}^2),
    \end{aligned}
\end{equation}
where $\mathbf{F}_\mathrm{D}^s$ is obtained by collecting the first $N_\mathrm{S}$ dominant right singular vectors of $\mathbf{H}_\mathrm{eff}^s \mathbf{F}(\mathbf{B}^s)$. The optimal $\mathbf{F}_{\mathrm{D},m}$ is thus obtained by extracting the first $N_\mathrm{S}$ dominant eigenvectors of $\mathbf{R}_m = \sum_{\mathbf{H}_{\mathrm{eff}}^s\in\mathcal{R}_m} \mathbf{F}_\mathrm{D}^s (\mathbf{F}_\mathrm{D}^s)^\mathsf{H}$.

\item\textbf{Loop back to step 4)} until convergence.
\end{enumerate}
\end{algorithm}


\section{Simulation Results}

In this section, we evaluate the performance of the proposed codebooks in limited feedback MIMO system. 

\subsection{Simulation Setup}
We consider a point-to-point MIMO system operating at $2.4$ GHz, where the transmitter uses reconfigurable pixel antennas and the receiver uses conventional antennas with fixed radiation patterns.
At the transmitter, each pixel antenna has $Q=39$ pixel ports and one antenna port. The impedance matrix $\mathbf{Z}$ and the open-circuit radiation pattern matrix $\mathbf{E}_\mathrm{oc}$ are obtained from the CST studio suite with $K=72$ spatial angle samples. 
The virtual channel $\mathbf{H}_\mathrm{V}$ follows an i.i.d. Rayleigh fading model. The noise power is normalized to $\sigma^2 = 1$.  

\subsection{Benchmark Schemes}

To comprehensively evaluate the performance of the proposed design, we involve the following three scenarios. 
\begin{enumerate}
    \item \textbf{Unconstrained MIMO Using Pixel Antennas}  assuming perfect CSI at the transmitter. In this scenario, the SEBO algorithm is deployed to search for the optimal antenna coder to maximize (\ref{eq:optimal_I}) in Section \ref{subsec:opt_tot}. This serves as an ideal performance upper bound. 
    \item \textbf{Limited Feedback MIMO Using Pixel Antennas} operating under finite feedback constraints. In this scenario, both the analog antenna coder and the digital precoder are selected from the offline-constructed finite codebooks ($\mathcal{B}$ and $\mathcal{F}_\mathrm{D}$) in Section \ref{sec:codebook} to maximize the spectral efficiency.
    \item \textbf{Unconstrained MIMO Using Conventional Antennas} assuming perfect CSI at the transmitter. In this scenario, the digital precoder is obtained using SVD on the channel combined with water-filling for power allocation. This serves as the performance lower bound without hardware reconfigurability.
\end{enumerate}
\begin{figure}[t]
    \centering
    \subfigure[Case 1 ($N=4, N_\mathrm{S}=4$).]{
        \includegraphics[width=0.732\linewidth]{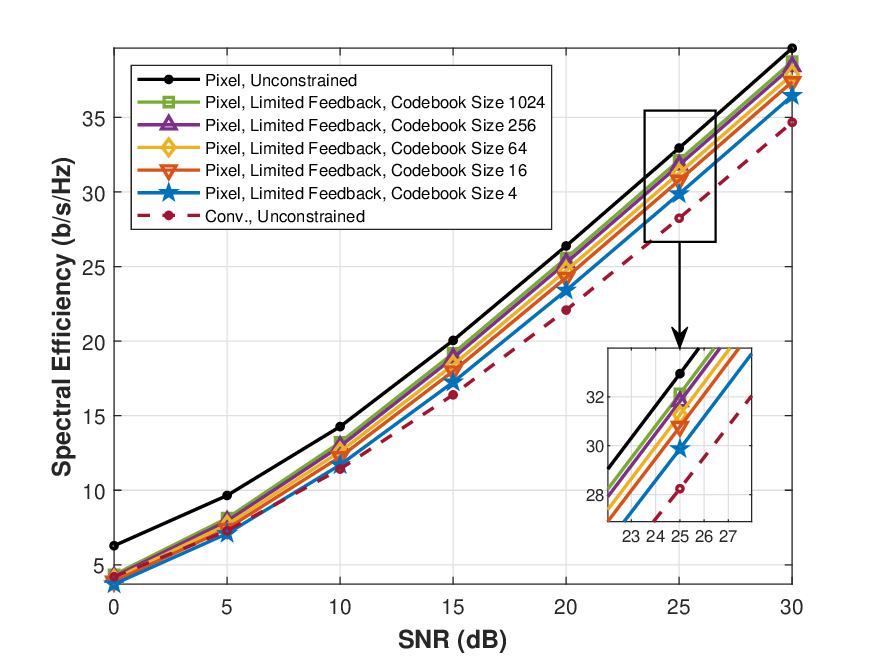} 
        \label{fig:case1_results}
    }
    \subfigure[Case 2 ($N=4, N_\mathrm{S}=3$).]{
        \includegraphics[width=0.732\linewidth]{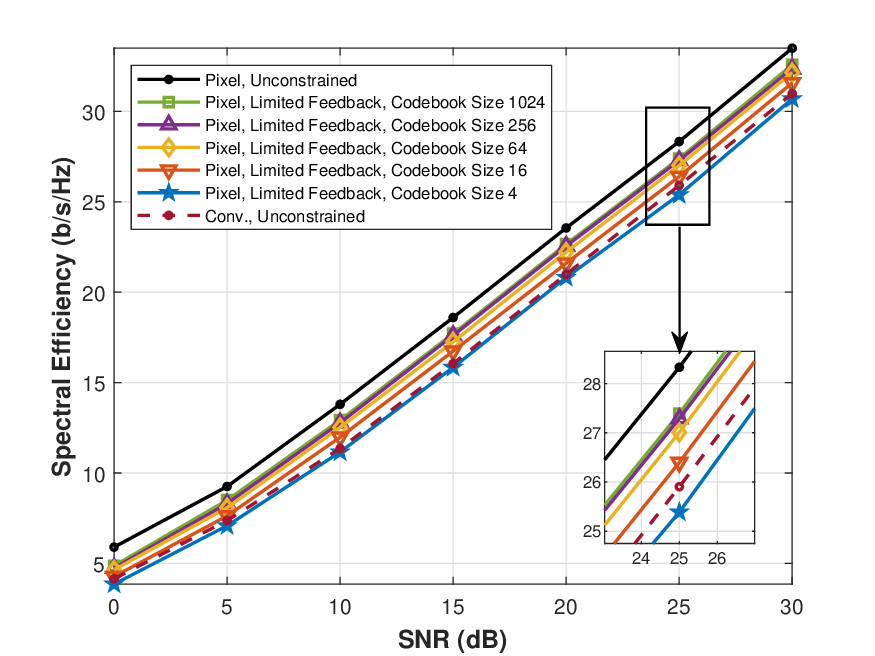} 
        \label{fig:case2_results}
    }
    \caption{Spectral efficiency versus SNR for different transmission setups. 
    In both cases, the antenna coder codebook and the digital precoder codebook are set to have the identical size $M$.}
    \label{fig:sim_results}
\end{figure}

\subsection{System Performance Analysis}

Figs. \ref{fig:case1_results} and \ref{fig:case2_results} illustrate the spectral efficiency as a function of signal-to-noise ratio (SNR) for the two cases discussed in Section \ref{sec:codebook}. In both cases, pixel antenna systems generally outperform conventional antenna systems due to additional degrees of freedom that pixel antenna provides to reshape the effective channel eigenmodes before digital processing. For each case, we have the following observations. 
    
In Case 1 ($N=N_\mathrm{S}$), limited feedback MIMO using pixel antennas with even a small codebook surpasses unconstrained MIMO using conventional antennas at high SNR. For instance, at SNR $=$ 25 dB, limited feedback MIMO using pixel antennas with $M=4$ can reach a spectral efficiency of 30 bps/Hz, which is 7\% higher than unconstrained MIMO using conventional antennas.

    In Case 2 ($N_\mathrm{S}<N$), limited feedback MIMO using pixel antennas does not always outperform unconstrained MIMO using conventional antennas due to the joint quantization of antenna coder and digital precoder. With a very low codebook size ($M=4$), the compound quantization loss in the analog and digital domains causes significant performance loss. This indicates that insufficient feedback bits may fail to highlight the benefit of pixel antennas. 
    However, for $M \ge 16$, the reconfigurability provided by pixel antennas successfully mitigates quantization losses. For example, at SNR $=$ 30 dB, limited feedback MIMO using pixel antennas with $M=64$ has spectral efficiency of around 32 bps/Hz, higher than 30.5 bps/Hz for unconstrained MIMO using conventional antennas. 

To sum up, while the proposed limited feedback MIMO using pixel antennas can effectively improve the system performance compared to systems using conventional antennas, it is important to carefully choose the codebook size according to specific transmission scenarios. 

\section{Conclusion}
In this paper, we addressed the critical challenge of overwhelming CSI acquisition overhead in pixel antenna-empowered MIMO systems by proposing a practical limited feedback framework. Specifically, we derived the optimal digital precoder under both total and unitary power constraints, establishing a theoretical foundation for spectral efficiency maximization.
We then developed a low-complexity offline codebook construction algorithm that effectively decouples the joint optimization of the antenna coder and digital precoder codebooks.
Simulation results confirmed that our proposed limited feedback scheme using pixel antennas outperforms unconstrained schemes using conventional antennas, demonstrating the benefits of using pixel antennas in easing the requirement of perfect CSI at the transmitter and enhancing system performance.

\bibliographystyle{IEEEtran}
\bibliography{refs}
\end{document}